**Title:** Evaluation of multiple imputation to address intended and unintended missing data in case-cohort studies with a binary endpoint.

**Authors:** Melissa Middleton[1,2], Cattram Nguyen[1,2], John B Carlin[1,2], Margarita Moreno-Betancur[1,2], Katherine J Lee[1,2]

1. Clinical Epidemiology & Biostatistics Unit, Murdoch Children's Research Institute, Melbourne Australia
2. Department of Paediatrics, The University of Melbourne, Parkville, Australia

**Corresponding Author:** Melissa Middleton (melissa.middleton@mcri.edu.au)

Murdoch Children's Research Institute, Royal Children's Hospital, 50 Flemington Rd, Parkville, VIC 3052

**Key words:** multiple imputation; inverse probability weighting; case-cohort; simulation study

**Abstract**

Case-cohort studies are conducted within cohort studies, wherein collection of exposure data is limited to a subset of the cohort, leading to a large proportion of missing data by design. Standard analysis uses inverse probability weighting (IPW) to address this intended missing data, but little research has been conducted into how best to perform analysis when there is also unintended missingness. Multiple imputation (MI) has become a default standard for handling unintended missingness, but when used in combination with IPW, the imputation model needs to take account of the weighting to ensure compatibility with the analysis model. Alternatively, MI could be used to handle both the intended and unintended missingness. While the performance of a solely MI approach has been investigated in the context of a case-cohort study with a time-to-event outcome, it is unclear how this approach performs with binary outcomes. We conducted a simulation study to assess and compare the performance of approaches using only MI, only IPW, and a combination of MI and IPW, for handling intended and unintended missingness in this setting. We also applied the approaches to a case study. Our results show that the combined approach is approximately unbiased for estimation of the exposure effect when the sample size is large, and was the least biased with small sample sizes, while MI-only or IPW-only exhibited larger biases in both sample size settings. These findings suggest that MI is the preferred approach to handle intended and unintended missing data in case-cohort studies with binary outcomes.

## BACKGROUND

The case-cohort study design is a powerful and cost-effective approach to study effects of exposures on outcomes, where the exposure may be costly to measure in cohort studies, for example metabolite levels[1]. In this study design, a subcohort is randomly selected from the main cohort and the expensive exposure information is only collected on the participants within the subcohort and on cases of the outcome, noting some subcohort members may also be cases. Hereinafter we refer to the subcohort and cases collectively as the study 'subset'. Analysis is then conducted on this subset, with the exposure intentionally missing 'by design' in the remainder of the cohort.

In such a design, it is important that the analysis accounts for the resulting unequal sampling probabilities due to all cases being selected into the subset (probability of selection = 1) and non-case subcohort members selected with a probability < 1[2]. Standard practice is to use inverse probability weighting (IPW) to account for this unequal sampling[3]. IPW involves discarding observations with missing exposure data (i.e. those not in the subset) and weighting the remaining observations in the analysis by the inverse probability of selection, to not only represent themselves, but also those not selected into the subset[4].

As with any study, it is common to have missing data due to non-response in several study variables (e.g. exposure and/or covariates). We will refer to this as unintended missing data. A popular approach to handling unintended missing data is multiple imputation (MI). MI is a two-stage process. In the first stage, imputed values are drawn from an approximate posterior distribution for the missing values dependent on the observed data[5]. Values are imputed several times to form *m* completed datasets. In the second stage, each completed dataset is analysed using the target analysis model and results are pooled across the *m* datasets using Rubin's rules to obtain an overall estimate for the parameter of interest with an estimated variance[6]. For MI to produce unbiased estimates with correct standard errors (SE), the imputation model needs to be compatible with the analysis model[7, 8]. Simply put, this means the imputation model should include all variables and features of the analysis model. In the context of case-cohort studies analysed using IPW, and weighted analyses more broadly, this means accounting for the weights used in the analysis model within the imputation model[9, 10].

There have been several approaches suggested in the literature for incorporating weights into the imputation model. One approach is simply to include the weights as a predictor in the imputation model. However, Carpenter et al. [9] demonstrated this approach can yield biased variance estimates, and instead proposed including the weights and their two-way interactions with all analysis variables in the imputation model. While the latter approach was found to perform well in terms of bias of point and variance estimates, the large number of parameters to be estimated may cause numerical problems for the imputation procedure. An alternative approach is to conduct stratum-specific imputation, whereby missing values are imputed separately within strata defined by similar weights [11, 12]. In the context of case-cohort studies with a binary endpoint, this is equivalent to conducting MI stratified by the outcome as all cases are assigned a weight of one and all non-cases a common weight greater than one. A drawback of stratum-specific imputation in case-cohort studies is the difficulty that may arise with small sample sizes within a stratum due to rare outcomes. An additional approach available in most statistical software is to use a weighted imputation model[11, 13]. Middleton et al. [12] evaluated the performance of these four approaches to handling missing covariate data in the context of case-cohort studies with a binary endpoint, and showed that the inclusion of the outcome, as a proxy for weight, as a predictor in the imputation model, without interaction terms, was an approximately unbiased and efficient approach to incorporate the analysis weights into the imputation model.

An alternative approach is to use MI to address both the intended and unintended missing data. The use of MI to handle intended missing data in case-cohort studies has previously been investigated in the context of a time-to-event outcome, where it was found to perform well provided the outcome and all variables in the analysis model were included in the imputation model[14-16]. However, these studies did not consider the scenario in which there are also unintended missing data. Keogh et al. [17] extended this work, comparing three approaches for using MI in a case-cohort setting with unintended missing data. They compared: the 'substudy' approach, which uses the subset only to fit an imputation model for unintended missing data and uses IPW to handle intended missing data; the 'intermediate' approach, which uses the full cohort to fit an imputation model for the unintended missing data, but limits the analysis to those within the subset and uses IPW to handle intended missing data; and the 'full' approach, which uses the full cohort for imputation of both intended and unintended missing data and conducts an (unweighted) analysis. They showed all approaches to have large gains in efficiency compared to a complete-case analysis (CCA), which conducts an unweighted analysis in participants with complete data only, with the full approach showing the largest gain. They did, however, find the intermediate approach to be more robust to misspecification of the imputation model than the full approach, which can be a concern when imputing the large proportion of intended missing information in case-cohort studies. A major limitation of the Keogh et al. [17] study was that they only considered the scenario where each variable could either have intended or unintended missing data, but not both, a scenario that is likely to arise in practice. It was also restricted to time-to-event analyses.

In the current study, we aimed to address these gaps by evaluating MI methods for handing both intended and unintended missing data in the exposure and/or confounders, in the context of a case-cohort analysis of a binary outcome. We considered the substudy, intermediate and full MI approaches, introduced by Keogh et al. [17], combined with the four approaches (i.e. including weights as a predictor, with and without interactions, stratum-specific imputation and weighted imputation) for incorporating weights into the imputation model, as well as an IPW-only and CCA (11 approaches in total since the full approach does not require weighting).

This paper is structured as follows. We first introduce a motiving example from the Barwon Infant Study (BIS), a birth cohort study in Victoria, Australia, and then describe the approaches for handling intended and unintended missingness in the case-cohort design that we compared. We then provide details of our simulation study, which was based on the motivating example and describe the application of the analysis approaches to the case study. We then present the results from the simulation and the case studies. We conclude with a discussion and recommendations for practice.

## METHODS

### Case Study

The motivating example for this manuscript comes from BIS, which is a population-derived birth cohort study of 1,074 infants born in the Barwon region of Victoria, Australia. The cohort profile and study design have been described elsewhere[18]. Due to the costly nature of biosample analysis, BIS has adopted the case-cohort design in several investigations of exposure effects on outcomes. The empirical investigation of interest here focusses on the association between vitamin D insufficiency (VDI) at birth, measured as 25(OH)$D_3$ serum metabolite levels below 50nM from cord blood, and the risk of food allergy at one-year, as determined by a combination of a positive skin prick test and a positive food challenge to one of five common allergens (sesame, peanut, cow's milk, egg and cashew)[19]. Of the infants who completed the one-year follow-up (n=894), all of the cases (n=61) and a random subcohort selected with a probability of 0.30 (n=324) were chosen for inclusion in the case-cohort study and had the exposure measured (noting some infants were in both). Of the 365 infants in the selected subset, VDI was only measured in 278 infants (76.2%), hence 23.8% of the subset had unintended missing data in the exposure.

The estimand of interest for the case study was the risk ratio (RR) for food allergy comparing those with VDI to those without. A standard outcome regression approach was used for its estimation, adjusted for family history of allergy (FamHx), "Caucasian" ethnicity (Eth), number of siblings (NSib), domestic pet ownership (PetOwn) and antenatal vitamin D supplement usage (AnteVD). Estimation used the modified Poisson regression approach with a logarithmic link and "robust" variance estimation, due to the known convergence issues with log-binomial regression[20]:

$$\begin{aligned}\log(\Pr(\text{FoodAllergy} = 1)) = {} & \theta_0 + \theta_1 \text{VDI} \\ & + \theta_2 \text{Eth} + \theta_3 \text{FamHx} + \theta_4 \text{PetOwn} + \theta_5 \text{AnteVD} \\ & + \theta_6 I[\text{NSib} = 1] + \theta_7 I[\text{NSib} = 2] \end{aligned} \tag{1}$$

Where $I[.]$ is an indicator function for the equality contained within the brackets (equal to 1 if the equality holds and 0 otherwise). The parameter of interest is $\log(\text{RR}) = \theta_1$. This is a slightly modified analysis to that used in the published version of this study, which used a log-binomial regression model to estimate the RR adjusted for a slightly different set of confounders. A description of the variables used for the current study can be found in Table 1, limited to participants with complete outcome data to align with the scope of this study.

**Analysis Methods to Account for the Missing Data**

Below we outline the approaches we considered for the handling of missing data in the analysis of case-cohort studies that have unintended missing data in the exposure and confounders. We comment on alternative approaches that we could have considered in the discussion.

**Table 1:** Description of variables in the Barwon Infant Study and their respective level of missingness in the full cohort and subset in participants with complete outcome data.

| Variable | Label | Full cohort (n=786) | | Subset (n=325) | |
|---|---|---|---|---|---|
| | | Summary | Missing (%) | Summary | Missing (%) |
| ***Outcome*** | | | | | |
| Food allergy at 1 year – *present – n(%)* | FoodAllergy | 61 (7.8) | 0 (0.0) | 61 (18.8) | 0 (0.0) |
| ***Exposure*** | | | | | |
| Vitamin D insufficiency at birth – *present – n(%)* | VDI | 132 (45.1) | 493 (62.7) | 109 (44.3) | 79 (24.3) |
| ***Confounders*** | | | | | |
| Ethnicity – *"Caucasian" – n(%)* | Eth | 573 (73.2) | 3 (0.4) | 240 (74.3) | 2 (0.6) |
| Domestic pet ownership – *present – n(%)* | PetOwn | 602 (77.4) | 8 (1.0) | 239 (74.2) | 3 (0.9) |
| Antenatal vitamin D usage – *present – n(%)* | AnteVD | 460 (79.3) | 206 (26.2) | 193 (76.0) | 71 (21.9) |
| History of family allergy – *present – n(%)* | FamHx | 675 (86.9) | 9 (1.2) | 284 (88.2) | 3 (0.9) |
| Number of siblings – *n(%)* | NSib | | 0 (0.0) | | 0 (0.0) |
| None | | 320 (40.7) | | 113 (34.8) | |
| One | | 281 (35.8) | | 130 (40.0) | |
| Two or more | | 185 (23.5) | | 82 (25.2) | |
| ***Auxiliary*** | | | | | |
| Maternal age at birth (*years*) – *mean(SD)* | MAge | 32.1 (4.8) | 3 (0.3) | 33.0 (4.3) | 0 (0.0) |
| SEIFA tertile – *n(%)* | SEIFA | | 14 (1.8) | | 6 (1.9) |
| Low | | 172 (22.3) | | 77 (24.1) | |
| Middle | | 148 (19.2) | | 61 (19.1) | |
| High | | 452 (58.6) | | 181 (56.7) | |

*SEIFA: Socioeconomic Index for Area*

### *MI-based Approaches*

We considered the three MI-based approaches as proposed by Keogh et al. [17]:

i. Subset (MI/IPW) – Subset data is used to fit an imputation model addressing the unintended missing data. The imputed datasets are analysed using a weighted regression model (i.e. to address the intended missing data), with the weights equal to the inverse probability of being selected into the subset

ii. Intermediate (MI/IPW) – The full cohort is used to fit an imputation model to address the unintended missing data. This approach involves also imputing the intended exposure missing data. However, the analysis is limited to observations within the subset only (i.e. non-subset imputed data is discarded) and a weighted analysis is performed on the subset, with the weights equal to the inverse probability of being selected into the subset

iii. Full (MI-only) – the full cohort is used to fit an imputation model imputing both the intended and unintended missing data, with an unweighted analysis performed on the full cohort (i.e. MI is used to handle both the intended and unintended missing data).

For all MI approaches, the imputation model included the outcome, exposure, confounders and two auxiliary variables unless specified otherwise. Auxiliary variables are variables which are included in the imputation model but not the analysis model and can improve the efficiency of MI if they are associated with the incomplete variables[21]. In this study, we have considered maternal age at birth and socioeconomic index for area (SEIFA) tertiles as the auxiliary variables. All incomplete variables (exposure and two confounders) were binary and were imputed from a logistic regression model within the MI framework of fully conditional specification[22] and 50 imputed datasets were generated.

Under the subset and intermediate approaches, the analysis model used IPW, with weights equal to the inverse probability of selection into the subset. The sampling weights for the *i*th observation, $w_{S,i}$ , are defined as:

$$w_{S,i} = \Pr(S_i = 1|Y_i)^{-1} \qquad (2)$$

where $S_i$ is an indicator for subset membership and $Y_i$ the outcome, for the *i*th individual.

Given all cases are included in the subset, the weights in expression (2) are 1 for cases. For non-case subcohort members, expression (2) is the inverse probability of subcohort membership for non-cases estimated by:

$$\widehat{w_{S,i}} = \left(\frac{m_0}{n_0}\right)^{-1} \quad (3)$$

where $n_0$ is the number of non-cases in the full cohort and $m_0$ the number of non-cases in the subcohort[23].

For both the subset and intermediate MI approaches, we considered four approaches to account for the unequal sampling weights in the MI procedure:

i. Weight only (*WO*) – incorporates the weights through inclusion of the outcome (as a proxy for the weight) as a predictor in each of the univariate imputation models within the FCS procedure.

ii. Weighted model (*WM*) – estimation of each of the univariate imputation models is weighted by the inverse probability of selection into the subset.

iii. Weight interactions (*WX*) – includes the outcome (as a proxy for the weight) as a predictor in each of the univariate imputation models, along with the two-way interactions between the outcome, and all analysis variables, excluding the variable being imputed.

iv. Stratum-specific imputation (*SS*) – the incomplete variables are imputed through an FCS procedure conducted separately within each weight class (i.e. cases and controls).

*IPW to handle intentional and unintentional missingness (Full-IPW)*

For completeness, we also considered a fully weighted approach. Here an IPW analysis was conducted on the complete records only, with weights representing the inverse probability of being a complete record, that is, records selected for the subset with complete data for all analysis variables.

The probability of being a complete record can be decomposed into the unintended and intended missingness probability components, assuming independence between the response indicator, $R_i$, and subcohort selection, $S_i$, given the outcome and the observed predictors of missingness (a plausible assumption), and independence between the observed predictors of missingness, $\boldsymbol{Z}_i$, and subcohort selection, $S_i$, given the outcome:

$$\Pr(R_i = 1 \,\&\, S_i = 1 \,|Y_i, \boldsymbol{Z_i}) = \Pr(R_i = 1|Y_i, \boldsymbol{Z_i}) \Pr(S_i = 1|Y_i) \qquad (4)$$

where $R_i$ is an indicator for having complete exposure and confounder data, and $\boldsymbol{Z}_i$ a set of completely observed predictors of (unintended) missingness that can include, but is not limited to, the analysis variables.

Under the *Full-IPW* approach, the probability of not having unintended missing data, $\Pr(R_i = 1|Y_i, \boldsymbol{Z_i})$, can be estimated by fitting a logistic regression model conditional on fully observed predictors of (unintended) missingness to the subset data.

A weight for the *i*th individual was then estimated by combining the sampling weight estimate and the inverse estimated probability of being a complete observation from the logistic model:

$$\widehat{w}_i = \widehat{w_{S,i}} \times \frac{1}{\widehat{\Pr(R_i = 1|Y_i, \boldsymbol{Z_i})}} \qquad (5)$$

*Complete case analysis (CCA)*

For comparison, a CCA was conducted, where observations with unintended missingness were deleted and IPW applied to the subset, using the sampling weights $\widehat{w_{S,i}}$, to address the intended missing data.

A summary of the analysis approaches is displayed in Table 2.

**Table 2:** Summary of analysis approaches used in the simulation and case studies

| Approach | Imputation Approach | Weights in analysis model | How missing data addressed | | Imputation sample | Analysis sample |
|---|---|---|---|---|---|---|
| | | | 'by design' | 'by chance' | | |
| CCA | None | sampling | IPW | Not addressed | N/A | Subset |
| Full IPW | None | combined* | IPW | IPW | N/A | Subset |
| MI-Sub-WO | FCS with outcome only | sampling | IPW | MI | Subset | Subset |
| MI-Sub-WM | FCS with weighted model | sampling | IPW | MI | Subset | Subset |
| MI-Sub-WX | FCS with outcome interactions | sampling | IPW | MI | Subset | Subset |
| MI-Sub-SS | Stratum specific FCS | sampling | IPW | MI | Subset | Subset |
| MI-Int-WO | FCS with outcome only | sampling | IPW | MI | Full cohort | Subset |
| MI-Int-WM | FCS with weighted model | sampling | IPW | MI | Full cohort | Subset |
| MI-Int-WX | FCS with outcome interactions | sampling | IPW | MI | Full cohort | Subset |
| MI-Int-SS | Stratum specific FCS | sampling | IPW | MI | Full cohort | Subset |
| MI-Full | FCS imputation | None | MI | MI | Full cohort | Full cohort |

**Combined weights are the product of inverse probability of selection into subcohort and inverse probability of being a complete observation*
*FCS: Fully conditional specification, IPW: inverse probability weighting, MI: multiple imputation*

## Simulation Study

A simulation study was conducted to compare the performance of the nine MI approaches for analysing case-cohort studies where there is unintended and intended missing data, along with the full IPW and a CCA approach, across a range of realistic scenarios. A complete-data analysis was also conducted, where an unweighted regression model was fitted to the simulated data prior to subcohort selection and missing data being induced, as a check of the data-generation process.

### *Data Generation Mechanisms*

Three scenarios were considered with respect to the full cohort sample size and the probability of subcohort selection. The first approximately replicates BIS, with a full cohort of 1,000 and a subcohort selection probability of 0.3. We also consider scenarios with a full cohort of 10,000, and a subcohort selection probability of either 0.1 or 0.2, mirroring the large sample sizes and smaller selection probabilities of other studies[14, 15].

Complete cohorts were first generated based on plausible causal relationships between the relevant variables and their missingness indicators as shown in Figure 1.

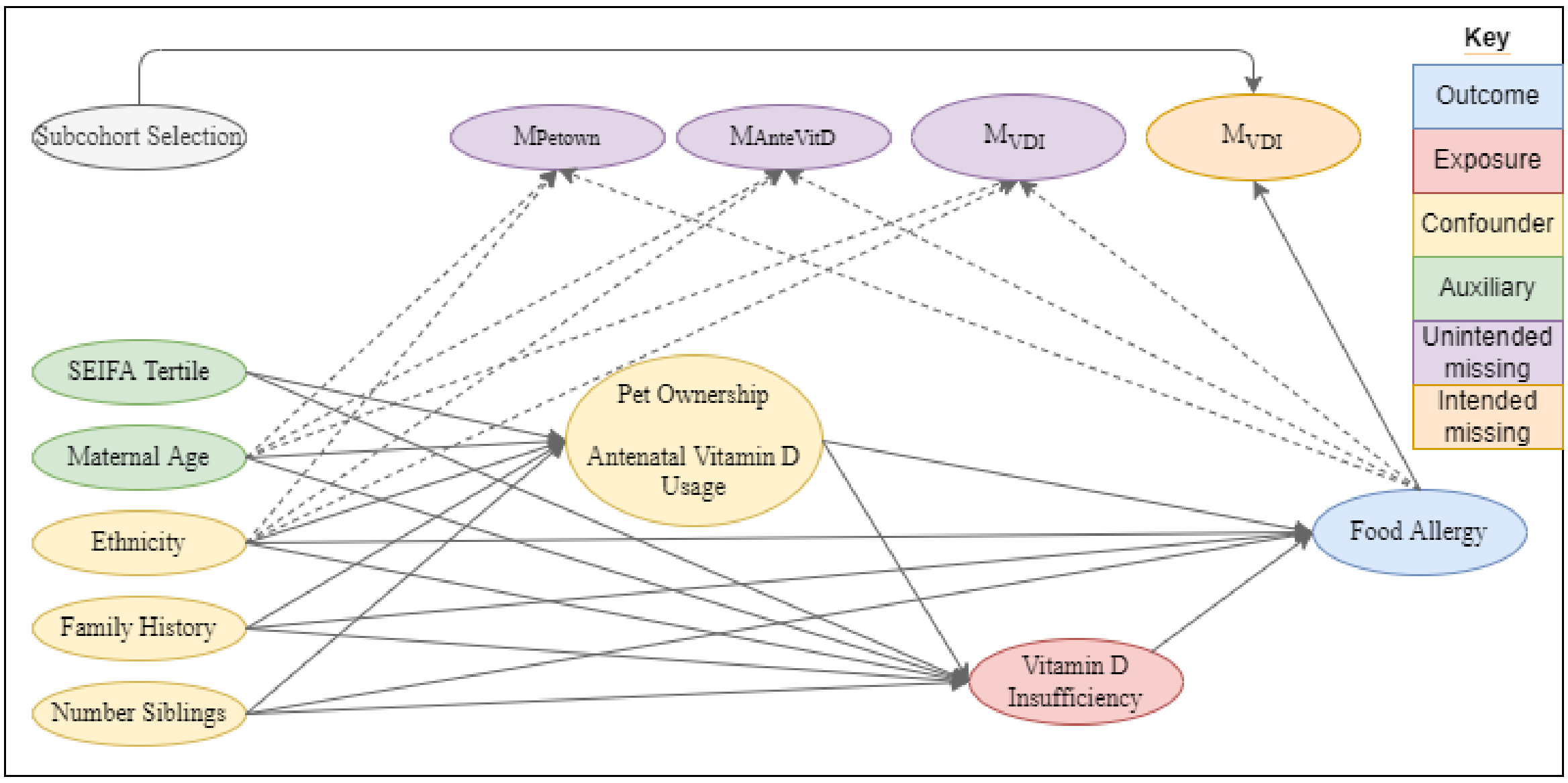

**Figure 1:** Missingness directed acyclic graph (m-DAG) depicting the assumed causal relationships between generated variables and their missingness indicators. Dashed lines represent associations present under the dependent missing data mechanism, but not the independent missing data mechanism.

The exposure, five confounders and two auxiliary variables were generated in a sequential manner using the models below:

i. Ethnicity

$$\text{Eth} \sim \text{Bernoulli}(p) \tag{6}$$

ii. Maternal age at birth

$$\text{MAge} = \delta_0 + \delta_1 \text{ Eth} + \epsilon \tag{7}$$

where $\epsilon \sim N(0, \sigma^2)$

iii. SEIFA tertile

$$\log\left(\Pr\left(\frac{\text{SEIFA} = 1}{\text{SEIFA} = 0}\right)\right) = \zeta_0 + \zeta_1 \text{ MAge} + \zeta_2 \text{ Eth} \tag{8}$$

$$\log\left(\Pr\left(\frac{\text{SEIFA} = 2}{\text{SEIFA} = 0}\right)\right) = \eta_0 + \eta_0 \text{ MAge} + \eta_0 \text{ Eth} \tag{9}$$

iv. History of family allergy

$$\text{logit}(\Pr(\text{FamHX} = 1)) = \iota_0 + \iota_1 \text{ Eth} \tag{10}$$

v. Number of siblings

$$\log\left(\Pr\left(\frac{\text{NSib}=1}{\text{NSib}=0}\right)\right) = \kappa_0 + \kappa_1\text{MAge} + \kappa_2\,\text{Eth} + \kappa_3\,I[\text{SEIFA}=1] + \kappa_4\,I[\text{SEIFA}=2] + \kappa_5\,\text{FamHx} \quad (11)$$

$$\log\left(\Pr\left(\frac{\text{NSib}=2}{\text{NSib}=0}\right)\right) = \lambda_0 + \lambda_1\,\text{MAge} + \lambda_2\,\text{Eth} + \lambda_3\,I[\text{SEIFA}=1] + \lambda_4\,I[\text{SEIFA}=2] + \lambda_5\,\text{FamHx} \quad (12)$$

vi. Domestic Pet Ownership

$$\text{logit}(\Pr(\text{PetOwn}=1)) = \rho_0 + \rho_1\,\text{MAge} + \rho_2\,\text{Eth} + \rho_3\,I[\text{SEIFA}=1] + \rho_4\,I[\text{SEIFA}=2] + \rho_5\,\text{FamHx} + \rho_5\,I[\text{NSib}=1] + \rho_6\,I[\text{NSib}=2] \quad (13)$$

vii. Antenatal vitamin D usage

$$\text{logit}(\Pr(\text{AnteVD}=1)) = \psi_0 + \psi_1\,\text{MAge} + \psi_2\,\text{Eth} + \psi_3\,I[\text{SEIFA}=1] + \psi_4\,I[\text{SEIFA}=2] + \psi_5\,\text{FamHx} + \psi_6\,I[\text{NSib}=1] + \psi_7\,I[\text{NSib}=2] \quad (14)$$

viii. Vitamin D insufficiency at birth,

$$\text{logit}(\Pr(\text{VDI}=1)) = \phi_0 + \phi_1\,\text{MAge} + \phi_2\,\text{Eth} + \phi_3\,I[\text{SEIFA}=1] + \phi_4\,I[\text{SEIFA}=2] + \phi_5\,\text{FamHx} + \phi_6\,I[\text{NSib}=1] + \phi_7\,I[\text{NSib}=2] + \phi_8\,\text{PetOwn} + \phi_9\,\text{AnteVD} \quad (15)$$

Finally, the outcome was generated per the target analysis model (1). We varied the strength of the exposure-outcome association, and the associations between the auxiliary variable, maternal age, and the incomplete variables, vitamin D insufficiency, antenatal vitamin D usage and pet ownership. Under 'observed' conditions, the associations were as estimated from the BIS case study, while under 'enhanced' conditions the exposure-outcome association was inflated to a RR of 2 (compared to RR=1.16 in BIS) and the associations of the exposure and missing confounders with the auxiliary variable maternal age were strengthened to represent an approximate 10-fold change in risk across the 30-year age range. An additional setting was considered, where the outcome generation model included an interaction between the exposure (VDI) and a confounder, ethnicity. This setting was designed such that the target analysis model was misspecified, as it excluded the interaction term, and enabled us to explore how the imputation models performed under a more complex but realistic scenario. The parameter values used for data generation under the various scenarios are given in the Supplementary Table S1.

Once the full cohort had been generated, unintended missing data were introduced into the two confounders, antenatal vitamin D usage and pet ownership, and the exposure. Two levels of missing data frequency were considered: low and high. In the low setting, missingness was generated such that 20% of records (in the full cohort) had at least one confounder missing and 10% had unintended missing data in the exposure, with an overall 25% of records having incomplete data. In the high missingness setting, each of these percentages was doubled.

Data were set to missing either using an independent missingness mechanism, where observations were randomly assigned to be missing with the desired proportions, or dependent on the outcome (expected to cause bias in the CCA), an auxiliary variable (expected to increase the efficiency of MI compared to CCA) and a confounder, as per Figure 1. The degree of dependency between the missingness indicators was varied to control the overall proportion of missing data, and the distribution of missing data patterns. Under the dependent missingness mechanism, data were set to missing based on the following models (with parameter values given in the Supplementary Table S1):

$$\text{logit}\left(\Pr\left(\text{M}_{\text{petown}} = 1\right)\right) = \nu_0 + \nu_1 \text{ FoodAllergy} + \nu_2 \text{ Eth} + \nu_3 \text{ MAge} \quad (16)$$

$$\text{logit}(\Pr(\text{M}_{\text{ante}} = 1)) = \tau_0 + \tau_1\ \text{FoodAllergy}\ + \tau_2\ \text{Eth} \\ + \tau_3\ \text{MAge} + \tau_4\ \text{M}_{\text{petown}} \quad (17)$$

$$\text{logit}(\Pr(\text{M}_{\text{vdi}} = 1)) = \omega_0 + \omega_1\ \text{Foodallergy}\ + \omega_2\ \text{Eth} \\ + \omega_3\ \text{MAge} + \omega_4\ I[\text{M}_{\text{petown}} = 1 \mid \text{M}_{\text{antevd}} = 1] \quad (18)$$

where $\text{M}_{\text{var}}$ is an indicator for missingness in variable "var".

The strength of associations in the dependent missingness mechanism were varied, with the 'observed' scenarios using estimates from BIS as values for the regression coefficients of substantive predictors in models (16)-(18), and the 'enhanced' scenarios doubling these coefficients. The values for $\nu_0$, $\tau_0$, $\tau_4$, $\omega_0$, and $\omega_4$ were iteratively chosen such that the desired proportions of missingness were achieved. All parameter values used are provided in the Supplementary Table S2.

Finally, the subcohort was randomly selected with the required probability of selection, and the exposure set to be intentionally missing in the non-subset members.

Altogether 26 scenarios were considered, comprised of 24 scenarios in a factorial design and an additional 2 scenarios where the interaction term was included in the data generation model. The factorial design consisted of: 2 levels of association strength between missingness indicators and causes, the exposure-outcome association, and the associations between the auxiliary variable, maternal age, and the exposure and incomplete variables; 2 types of missing data mechanism; 2 missing data proportions; and 3 sample sizes. The interaction term, and subsequent misspecification of the analysis model, was only considered in scenarios with a large sample size, high levels of dependent missing data and enhanced associations, to stress test the approaches. All scenarios are summarised in the Supplementary Table S3, along with a table summarising the variables across the simulation datasets for each scenario (Supplementary Table S4).

*Evaluation of Analysis Approaches*

Each simulated dataset was analysed using each of the approaches for handling missing data detailed previously to produce an estimate of the target parameter, the regression coefficient of the exposure in model (1): $\log(\mathrm{RR}) = \theta_1$.

Specifically, the performance of the nine MI, the IPW and the CCA approaches was evaluated in terms of the relative bias (percentage bias relative to the true value of the target parameter, $\theta_1$), empirical and model-based SEs, and the coverage of the 95% confidence interval (CI) for the target parameter. In calculating these measures in the scenario where the analysis model was correctly specified (outcome generated from model (1)), the true value of the parameter of interest was the coefficient for the exposure used during outcome generation ($\theta_1$). In scenarios where the analysis model was misspecified (outcome generated from a model including an exposure-confounder interaction), the true value was estimated as the average of the exposure coefficient estimates obtained when applying the target analysis model (model (1)) to 1,000 simulated populations of size 1,000,000. Monte Carlo standard errors (MCSE) are also reported.

A total of 2000 simulations were used for calculation of the performance measures, ensuring that the MCSE for a true coverage probability of 95% would be 0.49%[24]. Convergence issues were expected with the *WX* and *SS* methods in small sample sizes, as seen in previous work by the authors[12]. To ensure 2000 datasets in which all analysis approaches produced results, up to 3000 simulated datasets were generated, with datasets where the algorithm failed to converge for any of the approaches excluded from the calculation of performance measures, and convergence rates reported. All analyses were conducted in Stata 15.1[25].

**Implementation of Analysis Methods in the Case Study**

Each of the analysis approaches was applied to obtain estimates of the target parameter, $\theta_1$, in model (1) in the case study. To align with the simulation study, the analysis was limited to observations with complete outcome data (full cohort n = 786, subset n=325). The incomplete variables in the case study were: VDI (414/786 intentionally missing and 79/786 unintentionally missing), pet ownership (1% missing in full cohort, 0.9% missing in subset), antenatal vitamin D usage (26.2% missing in full cohort, 21.9% missing in subset), history of family allergy (1.2% missing in full cohort, 0.9% missing in subset), Ethnicity (0.4% in full, 0.6% in subset), SEIFA tertiles (1.8% in full, 1.9% in subset), and maternal age (<0.01% in full). Binary variables were imputed using a logistic regression model, categorical variables using an ordinal logistic regression model, and continuous variables using a linear regression model. When the analysis approach required the use of sampling weights, the weight for the non-cases in the subcohort, were estimated using the proportion of non-cases selected for exposure measurement, i.e. $(0.30)^{-1}$.

## RESULTS

### Simulation study

In all scenarios with a cohort size of 1,000, both the *WX* and both the *SS* MI approaches experienced convergence issues, with the convergence rate ranging from 81.9%-99.3% for the *SS* approaches and 84.5%-96.9% for the *WX* approaches. When the sample size was increased to 10,000, using enhanced associations, only the *WX* approaches experienced issues (99.5-99.9% convergence rate). All methods ran successfully in scenarios with a large sample size and observed associations. Full convergence results can be found in the Supplementary Figure S1.

Figure 2 displays the relative bias for all approaches and scenarios. In all scenarios with a small sample size and an independent missingness mechanism, most approaches were approximately unbiased (<5%), with the exception of *MI-Full* which showed some bias in scenarios with low levels of missing data (-8.7 – 6.5%), the *SS* approaches in scenarios with high levels of missing data (-6.4 – 6.0%), and the *WX* approaches which were biased in the scenario with high levels of missing data and enhanced associations (-5.8 – -6.1%). In all scenarios with a small sample size, dependent missingness mechanism, and an observed association, all approaches (including the complete data analysis) showed bias in the point estimate, with the largest biases for the *Full-MI* approach (-15.6 – 7.1%). In contrast, when there was a small sample size with enhanced associations and dependent missingness, the *Full-IPW*, subset and intermediate approaches were relatively unbiased (<5.3%), with slightly larger biases with the CCA and *Full-MI* approaches (6.3 – 8.1%).

In all scenarios with a large sample size and correct specification of the analysis model, all approaches were approximately unbiased (<5.9%), with the largest biases for the CCA for the dependent missingness scenarios as expected. When the analysis model was misspecified (i.e. omitted the interaction term of the data generating model), the CCA was biased (9.1 – 9.4%) with all other approaches approximately unbiased (-3.1 – 1.7%).

The *Full-IPW* and CCA approaches performed similarly in terms of the empirical SE, with all MI approaches exhibiting expected gains in precision (resulting in narrower CIs) across all scenarios (see Supplementary Figure S2). The relative error in estimating the SE for all methods and scenarios is presented in Figure 3 (see Supplementary Figure S3 for the estimated model-based SE). In scenarios with a small sample size, an independent missingness mechanism and enhanced associations, all MI methods using the *WX* and *SS* approaches showed considerable overestimation of the SE (relative error of 4% with 25% incomplete observations and 7-8% for 50% incomplete observations), which was not seen with the other approaches or other scenarios. When the sample size was 10,000 and the analysis model was correctly specified, *MI-Sub-WO* and *MI-Int-WO* tended to have larger relative error compared to the other MI approaches within the subset and intermediate analyses, with the exception of scenarios with 25% dependent missingness, where they performed comparably with an approximately unbiased estimation of the SE. In general, the CCA and the *Full-IPW* approaches performed similarly to each other, with negligible bias in SE estimation, across all scenarios. When the analysis model was incorrectly specified, all approaches were unbiased in the estimation of the SE when the probability of subcohort selection was 0.1. However, when the probability of subcohort selection was 0.2, MI-based approaches showed some under-estimation of the SE (ranging from -2.4 to -3.5%), compared to minimal underestimation for the *Full-IPW* (-0.3%) and CCA (-1.1%) approaches.

The coverage probability of the 95% CI is shown in Figure 4. Across all scenarios with correct specification of the analysis model the nominal coverage level of 95% was generally within the expected MCSE range for all approaches (93.9 – 96.2%), with the coverage probability closer to the expected probability of 95% as the sample size increased. When the analysis model was misspecified, the *Full-MI* approach and CCA showed under-coverage, ranging from 92.2% to 94.3%, while the subset and intermediate MI approaches and the *Full-IPW* approach showed close to the nominal coverage.

**Case study**

The estimated RR and its 95% CI obtained from applying each analysis method to the case study data are displayed in Figure 5. All methods produced similar point estimates. The *Full-MI* approach had a narrower CI compared to the subset and intermediate MI approaches, with all MI approaches having a narrower CI compared to the *CCA* and *Full-IPW* approaches.

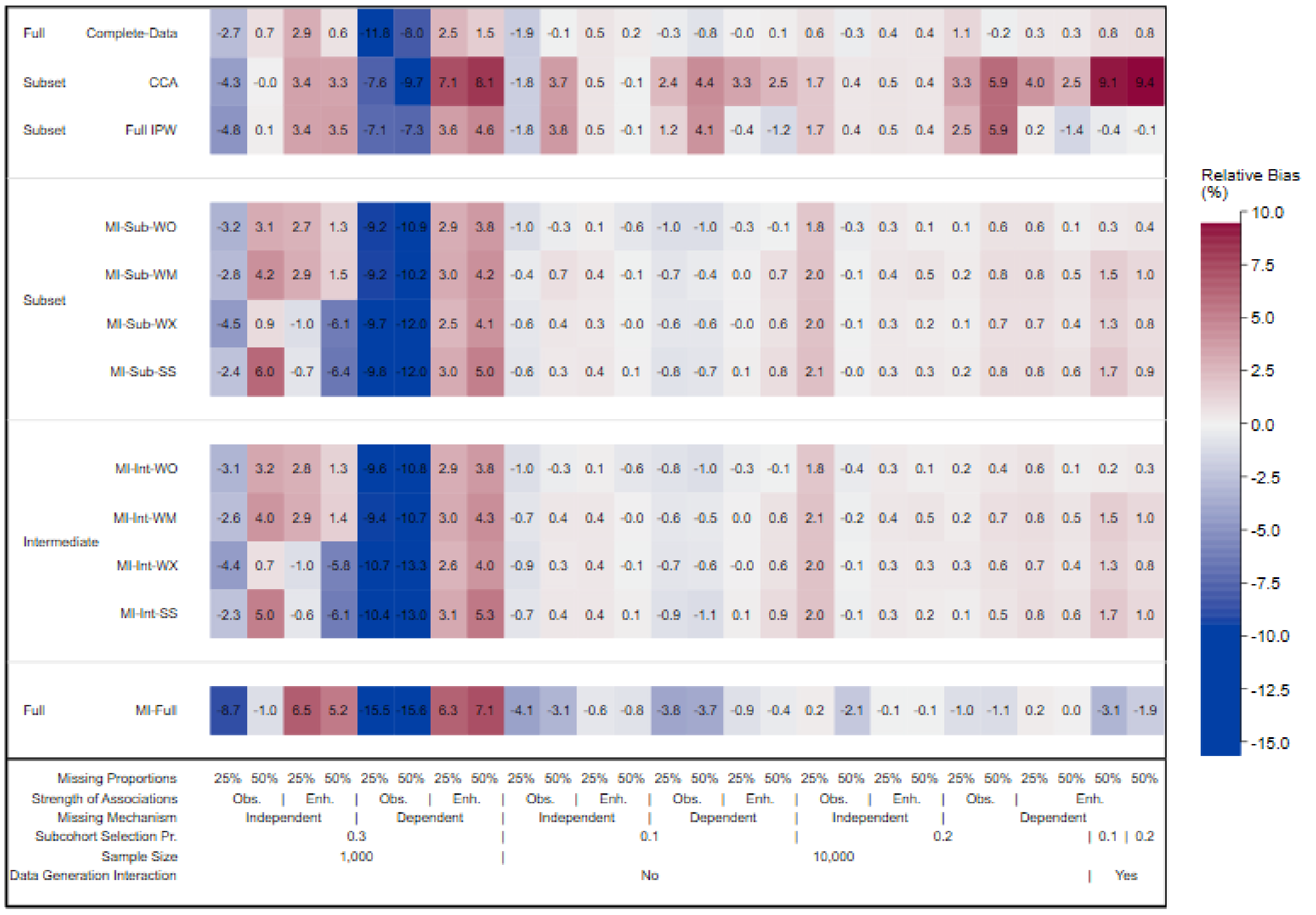

Monte Carlo Standards Errors range from .001 to .013, with mean .004

**Figure 2:** Relative bias (%) in the estimated coefficient for the target parameter across the 26 simulated scenarios

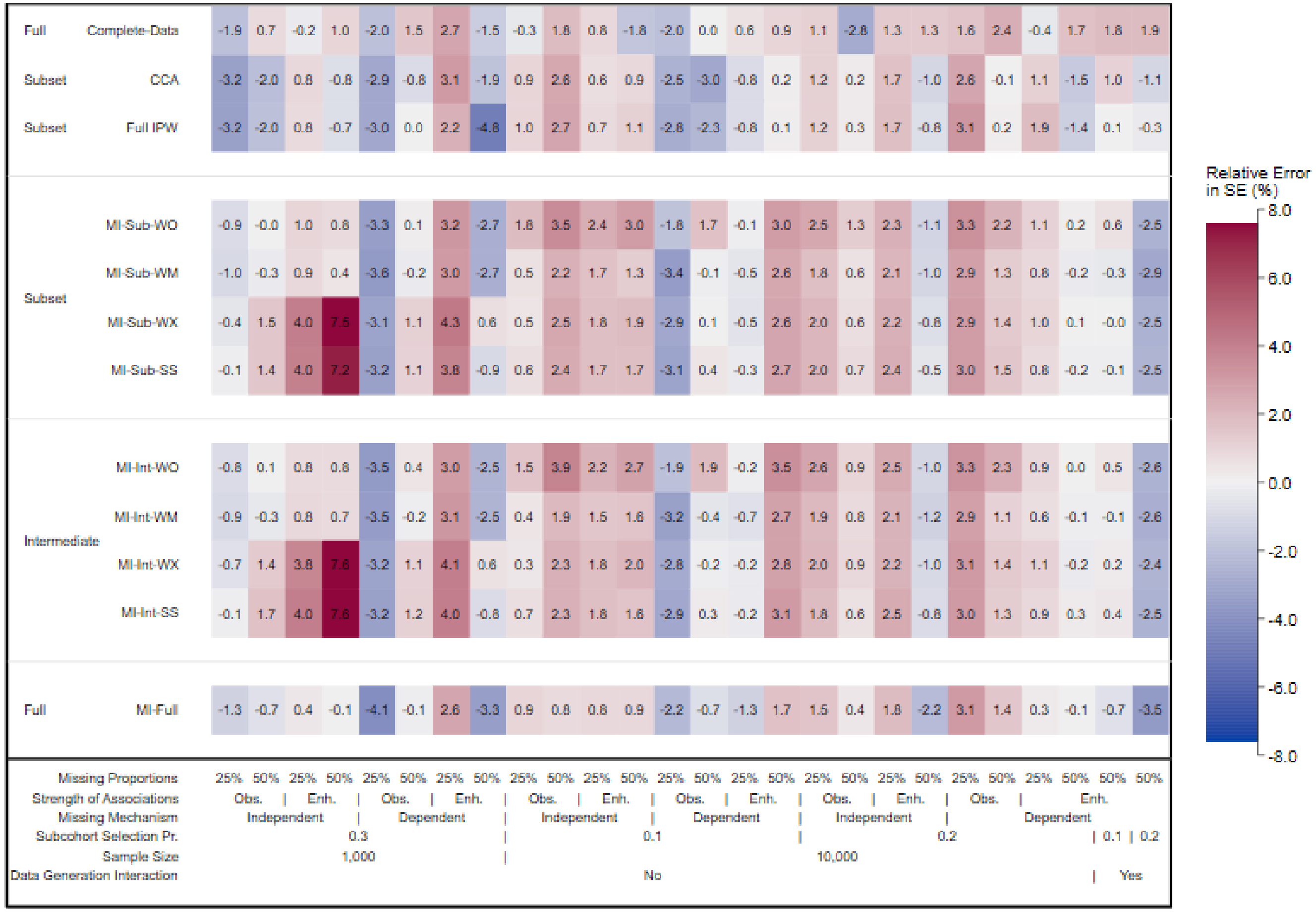

Monte Carlo Standards Errors range from 1.526 to 1.772, with mean 1.595

**Figure 3:** Relative error (%) in estimation of the standard error for the target parameter (comparison of empirical and model-based standard error) for each of the 26 simulated scenarios.

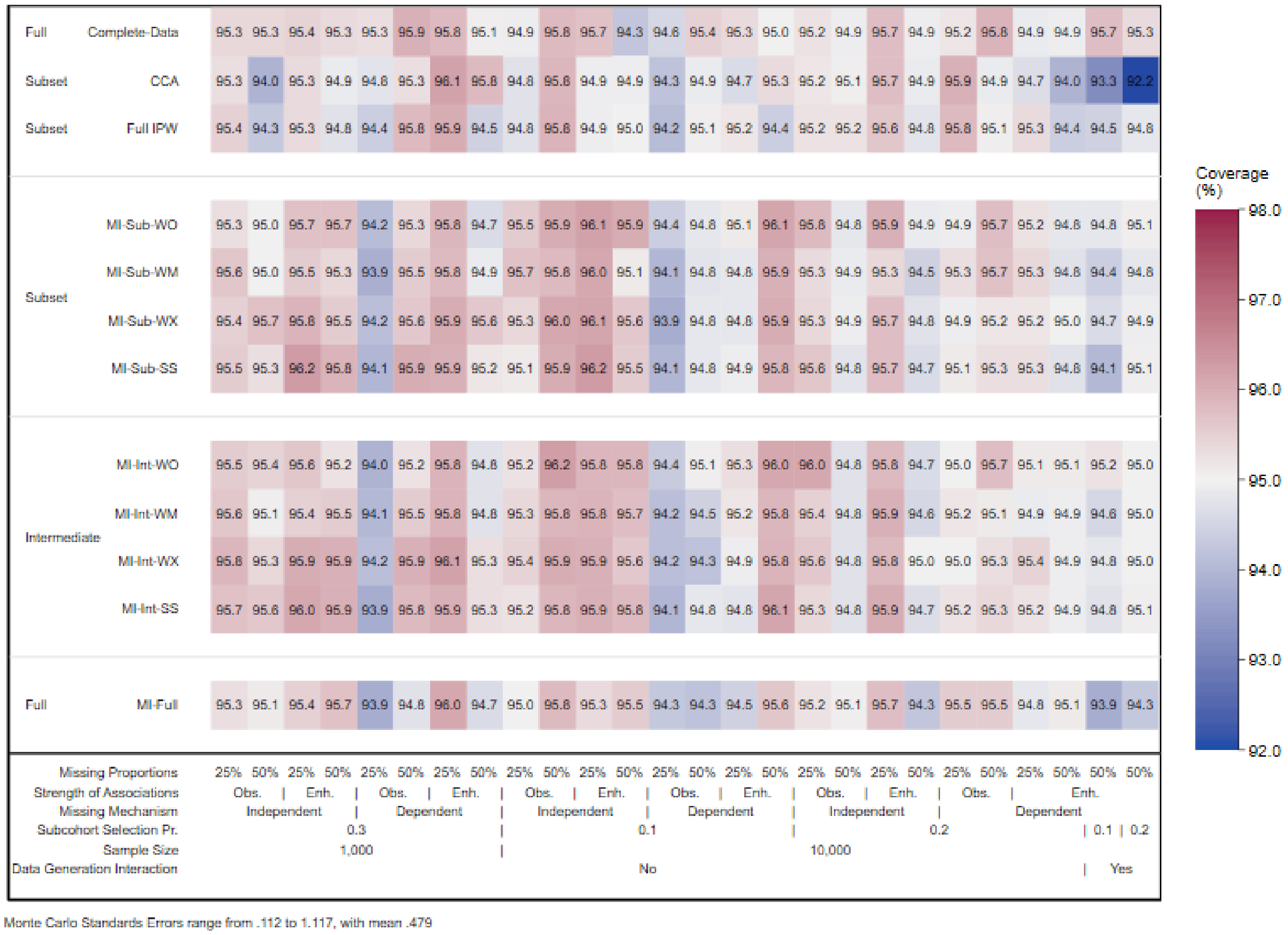

**Figure 4:** Coverage probability of the 95% confidence interval for each of the 26 simulated scenarios.

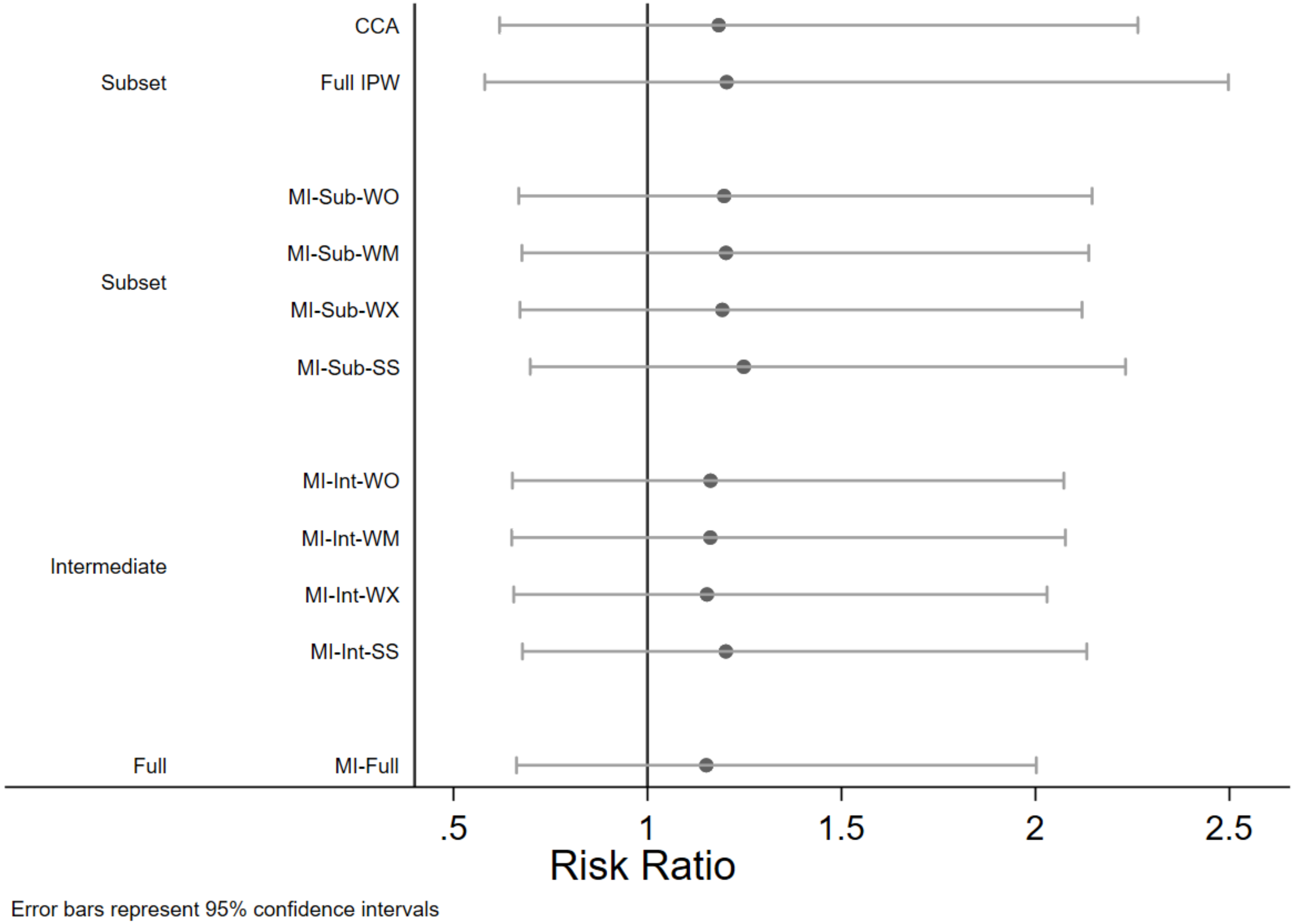

**Figure 5:** Estimated risk ratio and 95% confidence interval for the adjusted association between food allergy and vitamin D insufficiency estimated using the case study data

## DISCUSSION

This study aimed to evaluate approaches to handling intended and unintended missing data in case-cohort studies with a binary endpoint. We conducted a simulation study to compare the performance of 11 analytic approaches (nine MI approaches, a fully weighted approach and a CCA) across a range of scenarios.

When there was a small sample size, all analysis approaches, including the complete-data analysis, showed bias in the point estimate, which was not seen in scenarios with a large sample size. This is indicative of a finite sample bias in case-cohort studies, as previously demonstrated by Middleton et al. [12] and Keogh et al. [17]. While the approaches that used MI and IPW in combination (subset and intermediate MI approaches) generally performed similarly to the complete-data analysis in these small sample scenarios, larger biases were seen in the *Full-MI* approach that used MI to accommodate both the intended and unintended missing data. In these small sample size settings, combined MI/IPW with either the outcome and it's two-way interactions with analysis variables are incorporated into the imputation (*WX)* or the imputations are stratified by outcome status (*SS*), tended to result in slightly larger bias and overestimated SEs than including the outcome as a predictor (*WO*) or weighting the imputation model (*WM*).

In settings where there was a large sample size, the combined MI/IPW approaches that included the outcome as a predictor in the imputation model (*WO*) showed underestimation of the SE estimation (and narrower CIs) in some settings. However, this did not translate into under-coverage of the 95% CI, and therefore may not warrant concern in practice. In the analysis model misspecification settings, the approach using IPW alone to address both the intended and unintended missing data (*Full-IPW*) and the combined MI/IPW approaches which included the outcome as a predictor in the imputation model (*WO*) showed consistently lower biases for both the point estimate and SE. Overall, these results suggest that the combined MI/IPW approaches including the outcome as a predictor in the imputation model (*WO*) may be the preferred approach, with little difference between the subset and intermediate approaches.

Previous work had suggested the combined MI/IPW approach limited to the subset with the outcome included in the imputation model (*MI-Sub-WO),* performed well in handling confounders with unintended missing values in case-cohort studies with binary outcomes[12]. The results presented in the current simulation study suggest that the good performance of this approach extends to scenarios where the exposure is also unintended to be missing. While MI provided some expected gains in the precision of the exposure-outcome effect compared to the *Full-IPW* approach and a CCA, the simulation study results showed no apparent gain in bias or precision using a full or intermediate MI approach over the subset MI approach. These results are in contrast to those presented by Keogh et al. [17] who suggested an intermediate MI approach demonstrated greater gains in efficiency than a subset or full approach. It is important to note, however, that the subset approach may be subject to convergence issues in small case-cohort sample sizes, and an intermediate approach may be preferable in this setting. Interestingly, the *Full-MI* approach tended to show slightly larger biases compared to the subset and intermediate MI approaches, suggesting inferior performance of this approach.

While we considered four approaches for incorporating the weights into the imputation model, another possible approach is to include a subset indicator as a predictor, in the imputation model within the intermediate and full MI approaches, possibly including two-way interactions with all imputation variables. This approach would allow different associations between the predictors of missingness and the imputed variables within the subset (unintentional missing data only) and outside the subset (both intentional and unintentional missing data). We considered this approach in preliminary simulations but found that it had numerical problems due to the sparsity of the data and therefore it was omitted from our formal comparisons. Another approach that we considered was truncating the estimated probability of response used in the *Full-IPW* approach at the 5$^{th}$ and 95$^{th}$ percentile, to minimise the impact of highly variable weights. We found that this approach performed comparably to the untrimmed *Full-IPW* and hence was also omitted.

Our study was based on a realistic case-cohort setting and considered a large range of scenarios. While we have considered a small number of scenarios where the analysis model was misspecified, further exploration is needed to assess the appropriateness of MI in such settings. Due to limitations in the handling of missing outcome data in case-cohort studies using weighting approaches, given the weights are derived dependent on the outcome status, we have not considered missing outcome data in this study. This provides an avenue for future work. Another limitation is that we only considered IPW, MI and combined MI/IPW approaches. There are alternative analysis approaches, such as the semiparametric maximum likelihood and improved weighting approaches, as presented by Noma et al. [16], which could also be explored.

## CONCLUSIONS

Based on the findings in the current study, we do not recommend the use of a full MI approach to address both intentional and unintentional missing data in case-cohort studies with a binary endpoint, as this approach resulted in larger biases for estimation of the exposure-outcome association compared to a subset or intermediate approach. The subset and intermediate approaches performed similarly, with the inclusion of the outcome as a predictor in the imputation model showing the best performance across all scenarios, including those where the analysis model was misspecified. Therefore, we recommend addressing data missing unintentionally through MI applied to either the subset or full cohort with the inclusion of the outcome in the imputation model, along with all other analysis variables, and addressing intentional missing data through IPW (*MI-Sub-WO, MI-Int-WO)*.

## DECLARATIONS

### Ethics approval and consent to participate

The case study used data from the Barwon Infant Study, which has ethics approval from the Barwon Health Human Research and Ethics Committee (HREC 10/24). Participating parents provided informed consent and research methods followed national and international guidelines.

### Consent for publication

Not applicable

### Availability of data and materials

The datasets used and/or analysed during the current study are available from the corresponding author on reasonable request.

### Competing interests

The authors declare that they have no competing interests.

### Funding

This work was supported by the Australian National Health and Medical Research Council (Postgraduate Scholarship 1190921 to MM, career development fellowship 1127984 to KJL, investigator grant 2009572 to MMB and project grant 1166023). During part of the work MMB was also supported by an Australian Research Council Discovery Early Career Researcher Award (project number DE190101326) funded by the Australian Government. MM is funded by an Australian Government Research Training Program Scholarship. Research at the Murdoch Children's Research Institute is supported by the Victorian Government's Operational Infrastructure Support Program. The funding bodies do not have any role in the collection, analysis, interpretation or writing of the study.

### Authors' contributions

MM, CN, MMB, JBC and KJL conceived the project and designed the study. MM designed the simulation study and conducted the analysis, with input from co-authors, and drafted the manuscript. KJL, CN, MMB and JBC provided critical input to the manuscript. All of the co-authors read and approved the final version of this paper.

### Acknowledgements

The authors would like to thank the Melbourne Missing Data group and members of the Victorian Centre for Biostatistics for providing feedback in designing and interpreting the simulation study. We would also like to thank the BIS investigator group for providing access to the case-study data for illustrative purposes in this work.

# Supplementary material for manuscript titled: *Multiple imputation analysis of case-cohort studies with a binary endpoint*

## Contents

# Data generation procedure

Supplementary material provided in this section pertain to the data generation procedure of the simulation study as described in the manuscript.

Supplementary table S1 provides the exponentiated parameter values used to generate the complete data and the missing data indicators (1 if observation was incomplete for variable of interest and 0 is observation was complete). The table describes the model used for generation of each variable or indicator and provides the parameter value under each scenario. Parameter values under scenarios with observed association were chosen based on applying the data generation model to the case study data.

Supplementary table S2 provides the exponentiated parameter values for the parameters controlling the overall proportions of missing observations when modelling the missing indicator variables. Parameter values were chosen through an iterative process.

Supplementary table S3 provides a summary of the 26 scenarios considered in the simulation studies.

Supplementary table S4 provides summary measures for the 2,000 simulated datasets used for calculation of performance measures across the 26 scenarios. The outcome prevalence, case-cohort sample size (analysis sample used for subset and intermediate MI approaches), and of missing information are provided. The percentage of missing observations in the full cohort includes both the intentional and unintentional missing data and uses the sample size as the denominator. The percentage missing within the subset reflects the percentage of observations within unintentional missing data and uses the case-cohort sample size as the denominator. The first three letters of the scenario label reflect the strength of association, while the remaining letters reflect the missingness mechanism. Scenarios 1-3 have 25% incomplete observations, and scenarios 4-6 have 50% incomplete observations. Scenarios 1 and 4 have a cohort size of 1,000 and subcohort selection probability of 0.3, scenarios 2 and 5 have a cohort size of 10,000 and subcohort selection probability of 0.1, and scenarios 3 and 6 have a cohort size of 10,000 and a subcohort selection probability of 0.2. Scenarios ending in 'x' have an interaction term in the data generation model for the outcome.

**Supplementary Table S1:** Parameter values used in the generation of complete data and missing indicators, for observed and enhanced association scenarios.

| Dependent Variable | Model | Scenario | Intercept^ | *cauc* | *MAge* | *seifa=1* | *seifa=2* | *FamHx* | *NSib=1* | *NSib=2* | *PetOwn* | *AnteVD* | *VDI* | *VDI* × *cauc* | *FoodAllergy* |
|---|---|---|---|---|---|---|---|---|---|---|---|---|---|---|---|
| *Cauc* | Bernoulli | Both | 0.72 | | | | | | | | | | | | |
| *MAge* | Linear# | Both | 31.33 | 1.01 | | | | | | | | | | | |
| *SEIFA* | | | | | | | | | | | | | | | |
| *seifa=1* | Multinomial | Both | 0.10 | 1.65 | 1.05 | | | | | | | | | | |
| *seifa=2* | | Both | 0.17 | 1.17 | 1.08 | | | | | | | | | | |
| *FamHx* | Logistic | Both | 4.94 | 1.38 | | | | | | | | | | | |
| *NSib* | | | | | | | | | | | | | | | |
| *NSib=1* | Multinomial | Both | 0.02 | 0.81 | 1.11 | 0.74 | 0.92 | 2.29 | | | | | | | |
| *NSib=2* | | Both | 0.0004 | 0.77 | 1.21 | 1.43 | 1.11 | 3.39 | | | | | | | |
| *PetOwn* | Logistic | Obs Assoc.<br>Enh Assoc. | 11.84 | 1.17 | 0.97<br>0.90 | 1.33 | 0.74 | 0.83 | 1.11 | 1.36 | | | | | |
| *AnteVD* | Logistic | Both | 0.03 | 1.36 | 1.16 | 1.11 | 0.98 | 1.49 | 0.51 | 0.28 | | | | | |
| *VDI* | Logistic | Obs Assoc.<br>Enh Assoc. | 0.86 | 0.88 | 1.04<br>1.11 | 1.40 | 1.01 | 0.34 | 0.71 | 1.01 | 0.98 | 0.51 | | | |
| *FoodAllergy* | Poisson | Obs Assoc.<br>Enh Assoc.<br>Interaction | 0.07 | 1.08 | | | | 1.93 | 1.42 | 1.38 | 0.31 | 0.80 | 1.16<br>2.00<br>1.30 | <br><br>1.70 | |
| *M_petown** | Logistic | Obs Assoc.<br>Enh Assoc. | 0.57 | 0.66 | 0.89<br>0.90 | | | | | | | | | | 1.83 |
| *M_antevd** | Logistic | Obs Assoc.<br>Enh Assoc. | 1.84 | 0.71 | 0.96<br>0.90 | | | | | | | | | | 0.61 |
| *M_vdi** | Logistic | Obs Assoc.<br>Enh Assoc. | 0.77 | 0.93 | 0.98<br>0.90 | | | | | | | | | | 0.31 |

*^Probability given for Bernoulli models, intercept for linear models, base odds for Logistic models and base risk for Poisson models*
*#Error terms were generated with a mean of 0 and standard deviation of 4.75*
**Missing indicator models were additionally dependent on iteratively chosen values as shown in Table S2*

**Supplementary Table S2:** Iteratively chosen parameter values used to generate missing indicators for each data generation mechanism

| Parameter | Observed Assoc. | | Enhanced Assoc. | | Interaction Scenarios |
|---|---|---|---|---|---|
| | Low Missing | High Missing | Low Missing | High Missing | |
| $\nu_0$ | 2.14 | 3.08 | 1.79 | 2.78 | 2.78 |
| $\tau_0$ | -1.15 | -0.16 | 1.15 | 2.19 | 2.18 |
| $\tau_4$ | 3.31 | 2.35 | 3.54 | 2.54 | 2.55 |
| $\omega_0$ | -1.95 | -0.84 | 0.85 | 2.07 | 2.08 |
| $\omega_4$ | 1.56 | 0.47 | 1.69 | 0.44 | 0.42 |

**Supplementary Table S3:** Summary of the 26 scenarios considered in the simulation study.

| Scenario Label | Cohort sample size | | Strength of associations | | Outcome generation interaction | | Missing data mechanism | | Percentage unintended missingness | | Subcohort selection probability | | |
|---|---|---|---|---|---|---|---|---|---|---|---|---|---|
| | 1,000 | 10,000 | Observed | Enhanced | Present | Absent | Independent | Dependent | 25% | 50% | 0.3 | 0.1 | 0.2 |
| *Obsdep1* | X | | X | | | X | | X | X | | X | | |
| *Obsdep2* | | X | X | | | X | | X | X | | | X | |
| *Obsdep3* | | X | X | | | X | | X | X | | | | X |
| *Obsdep4* | X | | X | | | X | | X | | X | X | | |
| *Obsdep5* | | X | X | | | X | | X | | X | | X | |
| *Obsdep6* | | X | X | | | X | | X | | X | | | X |
| *Enhdep1* | X | | | X | | X | | X | X | | X | | |
| *Enhdep2* | | X | | X | | X | | X | X | | | X | |
| *Enhdep3* | | X | | X | | X | | X | X | | | | X |
| *Enhdep4* | X | | | X | | X | | X | | X | X | | |
| *Enhdep5* | | X | | X | | X | | X | | X | | X | |
| *Enhdep6* | | X | | X | | X | | X | | X | | | X |
| *Enhdep5x* | | X | | X | X | | | X | | X | | X | |
| *Enhdep6x* | | X | | X | X | | | X | | X | | | X |
| *Obsindep1* | X | | X | | | X | X | | X | | X | | |
| *Obsindep2* | | X | X | | | X | X | | X | | | X | |
| *Obsindep3* | | X | X | | | X | X | | X | | | | X |
| *Obsindep4* | X | | X | | | X | X | | | X | X | | |
| *Obsindep5* | | X | X | | | X | X | | | X | | X | |
| *Obsindep6* | | X | X | | | X | X | | | X | | | X |
| *Enhindep1* | X | | | X | | X | X | | X | | X | | |
| *Enhindep2* | | X | | X | | X | X | | X | | | X | |
| *Enhindep3* | | X | | X | | X | X | | X | | | | X |
| *Enhindep4* | X | | | X | | X | X | | | X | X | | |
| *Enhindep5* | | X | | X | | X | X | | | X | | X | |
| *Enhindep6* | | X | | X | | X | X | | | X | | | X |

**Supplementary Table S4:** Summary statistics for the 26 scenarios, calculated across the 2,000 simulated datasets.

| Scenario label | Outcome prevalence | | Case-cohort sample size | | Overall missing | | | |
|---|---|---|---|---|---|---|---|---|
| | | | | | Full cohort | | Subset | |
| *Obsdep1* | 6.6% | (0.76%) | 347 | (15) | 74.2% | (1.41%) | 25.4% | (2.34%) |
| *Obsdep2* | 6.6% | (0.25%) | 1597 | (37) | 88.2% | (0.33%) | 25.9% | (1.06%) |
| *Obsdep3* | 6.6% | (0.25%) | 2528 | (43) | 81.2% | (0.40%) | 25.6% | (0.88%) |
| *Obsdep4* | 6.7% | (0.76%) | 347 | (15) | 82.6% | (1.18%) | 50.0% | (2.58%) |
| *Obsdep5* | 6.6% | (0.24%) | 1596 | (38) | 92.0% | (0.27%) | 50.1% | (1.28%) |
| *Obsdep6* | 6.6% | (0.24%) | 2528 | (43) | 87.4% | (0.33%) | 50.1% | (1.00%) |
| | | | | | | | | |
| *Enhdep1* | 20.4% | (1.29%) | 443 | (16) | 67.5% | (1.47%) | 26.5% | (2.06%) |
| *Enhdep2* | 20.4% | (0.40%) | 2834 | (45) | 79.6% | (0.41%) | 28.0% | (0.86%) |
| *Enhdep3* | 20.4% | (0.40%) | 3631 | (49) | 73.5% | (0.43%) | 27.1% | (0.74%) |
| *Enhdep4* | 20.4% | (1.27%) | 443 | (16) | 78.2% | (1.31%) | 50.8% | (2.38%) |
| *Enhdep5* | 20.4% | (0.41%) | 2835 | (45) | 86.3% | (0.35%) | 51.6% | (0.95%) |
| *Enhdep6* | 20.4% | (0.41%) | 3632 | (48) | 82.2% | (0.38%) | 51.1% | (0.82%) |
| | | | | | | | | |
| *Enhdep5x* | 20.0% | (0.40%) | 2804 | (44) | 86.1% | (0.35%) | 50.4% | (0.96%) |
| *Enhdep6x* | 20.1% | (0.40%) | 3605 | (47) | 82.1% | (0.38%) | 50.3% | (0.85%) |
| | | | | | | | | |
| *Obsindep1* | 6.7% | (0.79%) | 347 | (15) | 74.0% | (1.36%) | 25.0% | (2.30%) |
| *Obsindep2* | 6.6% | (0.25%) | 1594 | (37) | 88.1% | (0.33%) | 25.0% | (1.15%) |
| *Obsindep3* | 6.6% | (0.26%) | 2530 | (44) | 81.0% | (0.39%) | 25.0% | (0.84%) |
| *Obsindep4* | 6.7% | (0.75%) | 347 | (15) | 82.6% | (1.20%) | 49.8% | (2.68%) |
| *Obsindep5* | 6.6% | (0.24%) | 1595 | (36) | 92.0% | (0.28%) | 50.0% | (1.30%) |
| *Obsindep6* | 6.6% | (0.25%) | 2529 | (43) | 87.4% | (0.32%) | 50.0% | (0.97%) |
| | | | | | | | | |
| *Enhindep1* | 20.3% | (1.25%) | 442 | (15) | 66.8% | (1.48%) | 25.0% | (2.06%) |
| *Enhindep2* | 20.4% | (0.42%) | 2833 | (46) | 78.8% | (0.41%) | 25.0% | (0.77%) |
| *Enhindep3* | 20.4% | (0.40%) | 3629 | (48) | 72.8% | (0.44%) | 25.0% | (0.70%) |
| *Enhindep4* | 20.4% | (1.25%) | 443 | (16) | 77.8% | (1.30%) | 49.9% | (2.35%) |
| *Enhindep5* | 20.4% | (0.41%) | 2834 | (46) | 85.8% | (0.35%) | 50.0% | (0.93%) |
| *Enhindep6* | 20.4% | (0.41%) | 3630 | (48) | 81.9% | (0.38%) | 50.0% | (0.82%) |

Mean and standard deviation provided for each summary statistics.

# Additional simulation study results

Supplementary material presented in this section provides additional results for the simulation study described in the manuscript.

Figure S1 provides the convergence rate (%) for each of the analytic approaches across the 26 scenarios. For scenarios with a cohort sample size of 1,000 and 25% of observations having unintentional missing data, the denominator was 2,320 simulations. For scenarios with a sample size of 1,000 and 50% of observations with unintentional missing data, the denominator was 3,000 simulations. For all remaining scenarios (cohort sample size of 10,000), the denominator was 2,056 simulations.

Figure S2 and S3 provide the empirical and model-based standard errors for the target parameter, respectively.

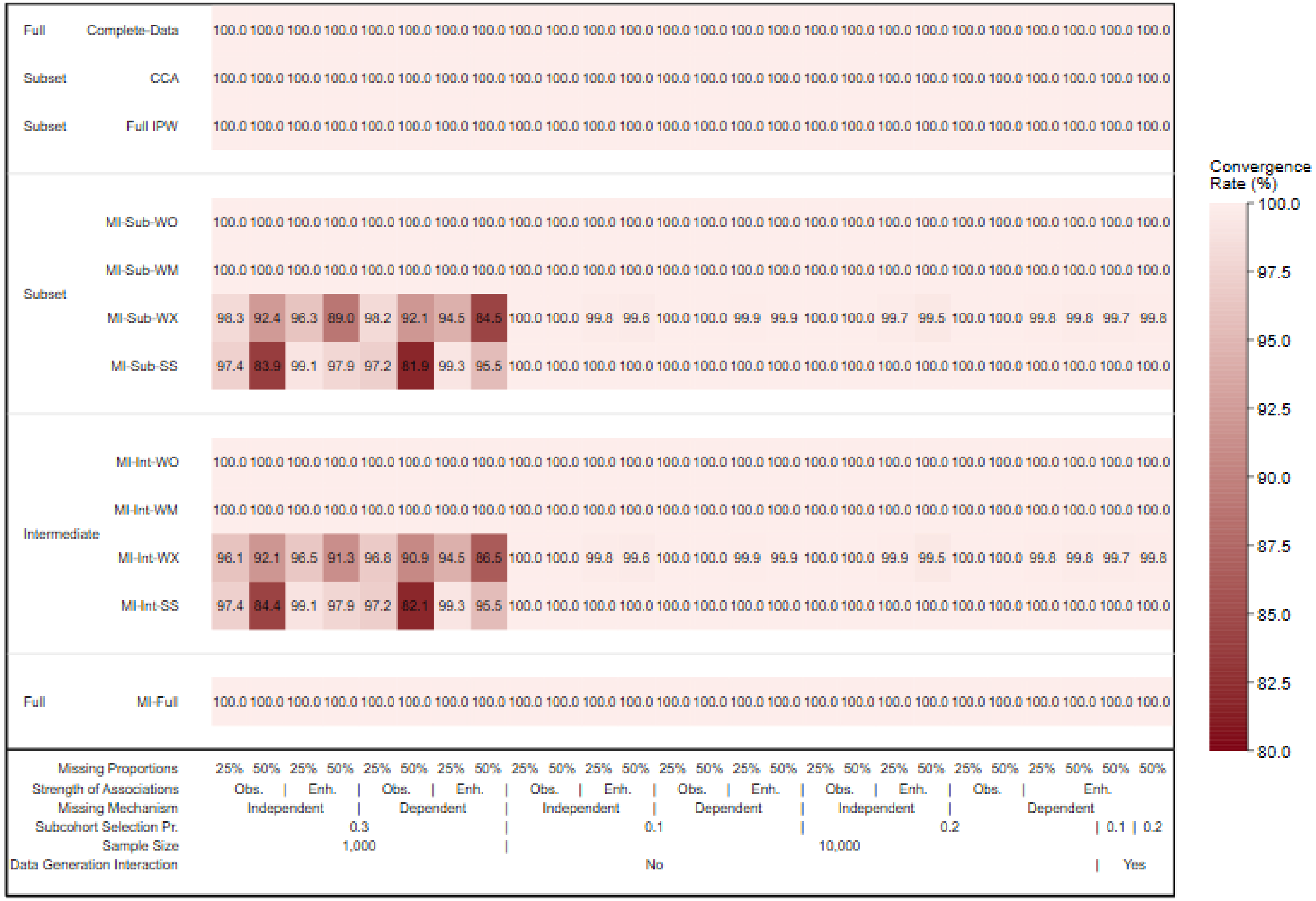

Convergence rate is calculated from: 2,320 simulations for scenarios with a sample size of 1,000 and 25% missing observations; 3,000 for a sample size of 1,000 and 50% missing; and 2,056 simulations for all other scenarios

**Figure S6:** Convergence rate for analytic approaches across the 26 scenarios

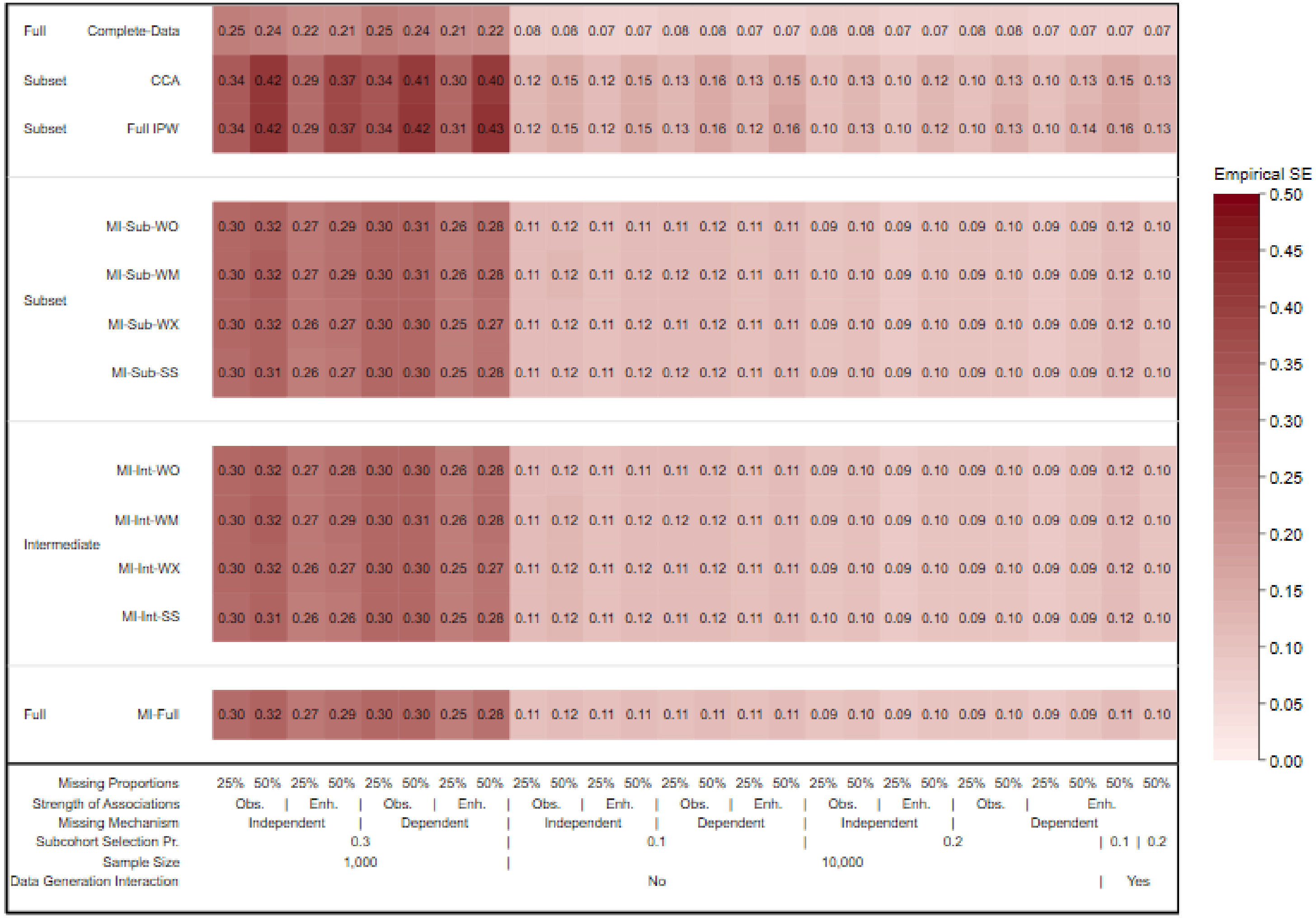

Monte Carlo Standards Errors range from .001 to .009, with mean .003

**Figure S7:** Empirical standard error for the target parameter value under each of the 15 analyses conducted, across the 26 scenarios

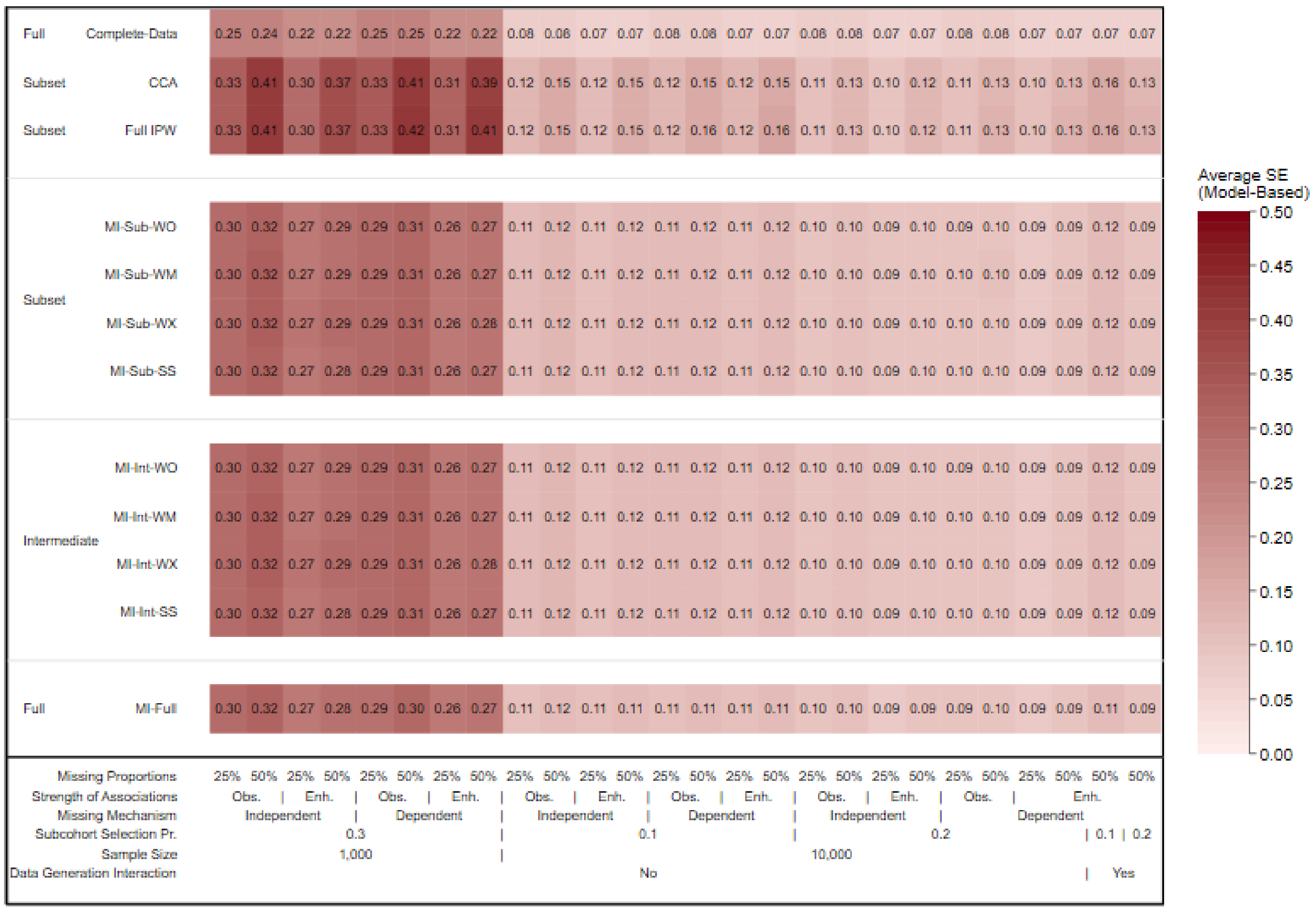

Monte Carlo Standards Errors range from 0 to .003, with mean 0

**Figure S8:** Model-based standard error for the analysis approaches across the 2,000 simulated datasets under each scenario